\documentstyle[12pt,epsf,aasms4]{article}
\newcommand{\ie }{{i.e.}}
\newcommand{\eg }{{e.g.}}
\newcommand{\etal }{{et~al.}}

\newcommand{\lya}{\mbox {Ly$\alpha$}}                   
\newcommand{\civ}{\mbox {C\,{\sc iv}}}                  
\newcommand{\heii}{\mbox {He\,{\sc ii}}}                
\newcommand{\mgii}{\mbox {Mg\,{\sc ii}}}                
\newcommand{\feii}{\mbox {Fe\,{\sc ii}}}                

\begin{document}

\begin{center}

{\Large{\bf On the absence of broad \mgii\ emission-line variability in
NGC~3516 during 1996.}}

\vskip 20pt

M.R.~Goad$^{1}$, A.P.~Koratkar$^{1}$, D.J.~Axon$^{2}$, 
K.T.~Korista$^{3}$, and P.T. O'Brien$^{4}$ \\

\vskip 20pt

{\small $^{1}$ STScI, 3700 San Martin Drive, Baltimore, MD 21218.}\\

{\small $^{2}$ Division of Physical Sciences, 
University of Hertfordshire, College Lane,
Hatfield, Herts, AL10 9AB, UK}\\

{\small $^{3}$ Dept. of Physics, Western Michigan University,
Kalamazoo, MI 49008.}\\

{\small $^{4}$ Dept. of Physics and Astronomy, University of
Leicester, University Road, Leicester, LE1 7RH, UK.}\\

\end{center}

\parindent=0.2in

\begin{center}

{\large{\bf Abstract}}

\end{center}

During the 1996 {\it HST/FOS\/} monitoring campaign of the Seyfert~1
galaxy NGC~3516, the UV continuum showed a factor of 5 variation;
correspondingly the strongest broad UV emission lines (\eg\ \lya,
\civ) displayed somewhat weaker, though still significant variations
(a factor of 2). In contrast, no variation was detected in the strong
broad \mgii\ $+$ UV \feii\ emission-line complex.  While historically
the flux in the broad \mgii\ emission line has varied (a factor of 2
within a few years), the profile shape has not been observed to change
over the past decade.  In contrast, the high ionization lines (HILs)
show both emission-line flux and profile changes on relatively short
timescales, which appear to be correlated with changes in the
continuum level. Significantly, on 1996, February 21, the profile
shapes of the HILs (\eg\ \lya, \civ) and low ionization lines (LILs,
\eg\ \mgii) were identical, at which time the continuum level was at
its highest.  These results impose stringent constraints upon viable
kinematic models of the BLR in NGC~3516, some of which we discuss
below.

Subject headings : galaxies : Seyfert --- galaxies : individual (NGC~3516) ---
galaxies : emission lines --- galaxies : kinematics and dynamics

\section{Introduction}

NGC~3516 is a Seyfert~1 galaxy with a well documented history of 
short-timescale, large-amplitude continuum and emission-line variability
(Koratkar \etal\ 1996). Historically, {\it IUE/SWP\/} spectra of
NGC~3516 have also displayed what has previously been referred to as a
strong, variable, broad absorption line (VAL,
FWHM~$\sim$200~km~s$^{-1}$) (Kolman et~al. 1993; Walter et~al. 1990;
Voit et~al. 1987), presumed to be similar to that observed in Broad
Absorption Line quasars (BALs, Weymann et~al. 1991).  The subsequent
disappearance of the supposed VAL between 1989 and the onset of the
1993 {\it IUE/SWP\/} monitoring campaign (Koratkar \etal\ 1996), may
be associated with the increased ionization state of the gas inferred
from the observed variation in the strength of the X-ray warm absorber
detected by {\it ROSAT\/} (Mathur et~al. 1997).

However, recent high resolution UV observations of the \civ\
emission-line region in NGC~3516 with the {\it HST/GHRS} (Crenshaw
et~al.  1998), show that NGC~3516 displays multiple intrinsic
absorption lines that are all blue-shifted and relatively narrow
(FWHM $\sim$500~km~s$^{-1}$), and which appear to remain invariant on
timescales of $\sim$~6~months, and are thought to originate in
the narrow-line region gas.  In light of these observations, Goad
et~al. (1999) argue that the supposed VAL is in fact multiple, narrow
absorption-line systems which {\it appear\/} broad at the lower
resolution of IUE. In this picture, the historical variation in the
strength of the VAL results from variations in the number of
narrow-line clouds coupled with changes in the shape of the underlying
broad emission-line.

In an ongoing effort to determine the nature of the broad
emission/absorption line regions and their long-term evolution, we
obtained 5 HST/FOS cycle~6 observations of NGC~3516, from 1995
December to 1996 November, covering the wavelength range
$\lambda\lambda$1150--3300~\AA.  In this paper we present the first
results of this study, highlighting the absence of emission-line flux
variations in the broad \mgii\ and UV \feii\ emission lines, and its
implications in terms of viable kinematic models of the BLR.

\subsection{Results from the 1996 HST/FOS monitoring campaign}

Goad et~al. (1999) showed that during the 1996 HST/FOS monitoring
campaign, NGC~3516 displayed the following continuum and emission-line
variability characteristics: (i) large amplitude, wavelength
dependent, continuum flux variations (a factor of 5 at $\lambda
1365$~\AA, c.f. a factor of 3 at $\lambda 2230$~\AA), (ii) correlated,
large amplitude, UV broad emission-line flux variations (a factor of 2
for both \lya\ and \civ), and (iii) the complete absence of variations
in both the strength and shape of the broad \mgii\ and UV \feii\
emission-line complex (Figure~1).

A comparison of the \mgii\ emission-line strength determined from
IUE/LWP observations taken in 1993 (Koratkar et~al. 1996) with the
present data, show that the flux in \mgii\ has apparently remained
constant for at least 4 years.  The absence of significant variability
in the \mgii\ emission-line strength, despite large amplitude
continuum and HIL flux variations, has also been
noted in lower resolution (pre--1989) IUE/LWP data (Koratkar
et~al. 1996) when the \mgii\ emission-line strength was a factor of 2
weaker. Interestingly, the observed doubling of the strength of the
broad \mgii\ emission-line between 1989--1993, appears to coincide
with the disappearance of the higher-velocity narrow absorption-line
components comprising the so-called VAL. It remains to be seen
whether these two phenomena are related. Significantly, despite the
notable increase in the \mgii\ emission-line strength between 1989 and
1993 its shape has remained unchanged for over a decade (Figure~2).
Moreover, in the present campaign, the shapes of the broad \lya, \civ\
and \mgii\ emission lines as observed on 1996, February 21, when the
continuum was in its {\em highest\/} state, are indistinguishable from
one another (Figure~3a, N.B. \lya\ is not shown).

\section{Implications for kinematic models of the BLR}

A viable kinematic model of the BLR in NGC~3516, must explain : (i)
the absence of both short-timescale ($\sim$ months) and long-timescale
($\sim$ a few years) variations in the \mgii\ emission-line flux and
shape, despite significant changes in the strength of the ionizing
continuum; and (ii) the remarkable similarity between the line shapes
of the HILs and LILs, at high continuum levels, when clearly these
lines form under very different physical conditions.

\subsection{Flux variability constraints}

An absence of response in the \mgii\ emission line can result for a
variety of physical reasons~: (i) the continuum band driving the line
is invariant; (ii) the emission-line is insensitive to continuum
variations; and (iii) the line emitting region is physically extended.

Here we address the shortcomings of each of these scenarios.

(i) While it is true that \mgii\ and \lya\ respond to very different
continuum bands (\mgii\ is mostly sensitive to the continuum from
600--800~eV (Krolik and Kallman 1988) whereas \lya\ is driven mainly
by Lyman continuum photons), simultaneous multi-wavelength
observations of a small number of AGN in the UV, optical and soft
X-ray band-passes, indicate that the amplitude of the continuum
variations generally increase toward shorter wavelengths (Romano and
Peterson 1998).  Although we cannot confirm whether such a trend holds
at 600--800~eV, the Seyfert~1 galaxies for which simultaneous
observations were made in the UV/EUV, NGC~5548 (Marshall et~al. 1997)
and NGC~4051 (Cagnoni et~al. 1998), showed increased variability at
shorter wavelengths. Thus, the evidence suggests that the lack of
\mgii\ and \feii\ variations is not due to an absence of continuum variability in
the bandpass responsible for driving these lines.

(ii) Goad et~al. (1993, 1995) and O'Brien et~al. (1995) demonstrated
that the line responsivity $\eta$, the fractional change in line
emissivity for a given fractional change in the continuum level, is
relatively modest for \mgii\ ($\eta\sim0.2-0.3$), when compared to the
approximately linear response ($\eta \approx 1$) of the HILs.  This
differences arises because for a single cloud and a fixed continuum
shape, the variation in the position of the hydrogen ionization front
is generally small compared to the overall extent of the partially
ionized zone where LILs such as \mgii\ are formed.  For a fixed
continuum shape this condition holds true over a large range in
ionizing continuum luminosity (factors of 10 or more). Although the
\mgii\ line responsivity is small, given the factor of 2 change in the
\civ\ emission-line flux, we should have detected $\sim 40$\%
variation in the \mgii\ emission-line flux. Since no significant
($<$7\%) variation was detected, low emission-line responsivity cannot
be the {\em sole\/} explanation for the absence of \mgii\
emission-line variations.

(iii) Finally, the lack of response in the \mgii\ emission line can
also result if the \mgii\ line-emitting region is physically large,
and has thus yet to respond to the observed continuum variations.
While plausible, given the short-timescale response of the \civ\
emission-line to the continuum variations ($\sim$ 4.5 days, Koratkar
et~al. 1996), it is difficult in this picture to explain the
similarity in shape of the \civ\ and
\mgii\ emission-line profiles at high continuum levels, particularly
if the size of the regions where these lines originate differ by a few
orders of magnitude.

The similar ionization potential of \feii\ and \mgii\ suggests that
these lines are formed in similar regions, hence it is unsurprising
that the \feii\ emission lines display a similar lack of response to
continuum variations (N.B. relatively few of the \feii\ lines
(although strong) are resonance lines, although others are pumped by
the continuum, especially in the presence of extra thermal particle
speeds).  The smoothness of the \feii\ complex in NGC~3516 when
compared to similar observations of I~Zw~I (Goad et~al. 1999), a
quasar with strong narrow \feii\ emission-lines (Laor
\etal\ 1997), supports this finding, and suggests that the \feii\
lines in NGC~3516 are indeed produced in high velocity gas (FWHM$\sim
4500$~km~s$^{-1}$).

\section{Profile variability constraints}

The similarity in shape between the HILs and LILs at the highest
continuum levels suggests that in the high-state both \civ\ and \mgii\
arise in kinematically similar regions.  However, if the lack of
response of \mgii\ is not the result of an absence of variability in
the continuum band driving this line, then the difference in response
timescale between the HILs and \mgii\ suggest that these lines {\it
do\/} arise in spatially distinct regions, with \mgii\ formed
in a region which extends over several light-months or more.
Figure~3b indicates that between the low and high-states, the core of
the \civ\ emission line shows a deficit of response redward of line center from $0$ to 
$+$4000~km~s$^{-1}$ (even after allowing for possible contamination by the variable
absorption-line and narrow emission-line components) and an enhanced
response in the far red-wing ($>5000$~km~s$^{-1}$). This may result from : (i)
reverberation within a radial flow; (ii) a change in the radiation
pattern of the ionizing continuum; or (iii) a change in the spatial
distribution of the line emitting gas.

Taken together, this
evidence places severe constraints upon the spatial distribution and
kinematics of the line-emitting gas in NGC~3516. For example, based on
this data the following models can be excluded :

\begin{itemize}

\item[(i)]{Radial flows at constant velocity, 
illuminated by either an isotropic or anisotropic continuum source
(Goad and Knigge 1999).}

\item[(ii)]{Radial flows in which \mgii\ and \civ\ arise in
azimuthally separated cloud populations, either through spatial
segregation or anisotropic illumination.}

\item[(iii)]{A spherical distribution of clouds in Keplerian motion
and illuminated by an isotropic continuum source.}

\end{itemize}

We propose two models which may account for the observed emission-line
variations: (i) an anisotropic continuum source model and (ii) a
terminal wind model (i.e. a flow that accelerates from rest until it
reaches constant velocity).

(i) The anisotropic continuum model of Wanders et~al. (1995) can
broadly reproduce the gross details of the observations reported here.
In this model, the continuum is assumed to be comprised of two
components, an anisotropic variable component responsible for driving
the variations in the HILs and an isotropic non-variable component
responsible for producing the \mgii\ and \feii\ emission.  To match
the observed variations, the BLR must be comprised of two distinct
cloud populations, a low column density population producing
significant HIL emission, and a spatially distinct high-column density
population producing both HILs and LILs. The variable ionizing
continuum component preferentially illuminates the low-column density
population, giving rise to the variations in \lya\ and \civ\
emission-lines, whereas the LILs arise predominantly from the
high-column density clouds illuminated by the non-variable isotropic
continuum component.  If the ionization parameter $U$ is large enough,
and the EUV is not absorbed, these high column density clouds could
also produce significant HIL emission. To account for the similarity
in profile shape in the high-state, we further assume that the
variable ionizing continuum component becomes more isotropic when
brighter, so that the \civ\ emission arises predominantly from the
high-column density clouds. Since the FWHM of the low and high-state
\civ\ emission-line profile is effectively the same, the velocity
field cannot be predominantly radial, otherwise a different range in
projected velocity for the HILs and LILs will result. Instead a
randomized or chaotic flow is favored.  While plausible, this model
requires a high-degree of fine tuning to produce the detailed profile
changes observed in the HILs. Hardest to explain are the reversals in
HIL profile shape, from red-asymmetric at low continuum levels to
blue-asymmetric at high continuum levels (Figure~3b).

(ii) One possible scenario which may account for the above
observations is a terminal wind model for the BLR\footnote{The best
available variability data on NGC~3516 (Wanders et~al. 1993; Koratkar
et~al. 1996) suggest that the BLR kinematics are not dominated by
radial motion, but may be consistent with a mixture of velocity
components.}.  A terminal wind is the one physical structure which can
display the same range in projected velocities on both small and large
spatial scales.  The wind may be spherical or non-spherical.  For a
non-spherical wind, for example a bi-conical flow, similar profiles
for the HILs and LILs will result {\it regardless\/} of their
respective radial emissivity distributions, provided that the opening
angle of the cone remains constant. However, to produce profiles
similar to that observed here (\ie\ broad wings and narrow cores), we
require a strong azimuthal dependence on the velocity field, or a
significant circularized velocity component to the flow (Goad and
Knigge 1999).  That is, the simplest terminal flow model in which the
velocity field is constant everywhere cannot produce the correct
profile shapes. However, hydromagnetic wind models (\eg\ Emmering
et~al. 1992; Bottorff et~al. 1997) do exhibit these basic properties.
While the emission-line flux variations of the HILs are assumed to be
driven by continuum variations, we propose that in this model the
changes in profile shape at low continuum levels are in part due to
changes in the formation radius of the HILs.
Specifically, we propose that in the low-state, conditions within the BLR gas
are such that a large fraction of the \civ\ emission arises at the base of the
wind, in the region where the gas begins to accelerate (Goad and Knigge 1999).

\section{Conclusions}

The picture we envisage is one in which the steady-state profile is
dominated by the dynamics of the system, with lines formed in
kinematically similar regions, possibly a terminal flow.  The
stability of the \mgii\ profile on timescales of several years,
provide strong supporting evidence for the existence of such a
structure.  Moreover, the absence of significant flux variations in
the \mgii\ line on timescales of a few years suggest that this region
is physically extended in size. The historic variation in the \mgii\
emission-line flux, but not its profile, could be the result of an
accretion event, a disk instability, or an increase in the wind
density and may be linked to the disappearance of the \civ\ VAL.

\begin{center}
{\bf Acknowledgements}
\end{center}

The results presented here are based upon observations made with the
NASA/ESA Hubble Space Telescope, obtained at the Space Telescope
Science Institute, which is operated by the Associated Universities
for Research in Astronomy, Inc, under NASA contract NAS5-26555. MRG
and AK acknowledge research grant GO-6108 for financial support. We
would also like to thank the referee for many helpful suggestions.
MRG would also like to thank Christian Knigge for many useful
discussions concerning disk-wind modeling.

\newpage

\begin{center}
{\bf Figures}
\end{center}

\figcaption[]{The high-state (epoch 2, solid line) and low-state
(epoch 5, dotted line) continuum subtracted \mgii+\feii\ emission-line
complex and their difference (dashed line). For clarity, each spectrum
has been binned into 2~\AA\ wide bins.  The derived errors (poissonian)
in the difference spectrum, displaced for clarity, are indicated by
the vertical bars. We detect no significant variation in the broad
\mgii\ and UV \feii\ emission-line complex. The absorption lines are Galactic 
in origin.}

\figcaption[]{A comparison between the normalized 1996 HST/FOS \mgii\
emission-line profile (smoothed to the resolution of the IUE/LWP),
with that determined from a) the 1993 IUE/LWP monitoring campaign, and
b) the IUE/LWP archival data.  Note that for the upper panel no
scaling between the HST and IUE spectra has been applied, the
emission-line fluxes and profile shapes are the same.  The solid line
in the lower panel shows the old average IUE/LWP \mgii\ flux scaled
relative to the flux measured with HST.  While the \mgii\
emission-line flux clearly varies (a factor of 2.3), the profile shape
remains invariant on timescales of several years. The absorption dip
blueward of line centre is due to Galactic \mgii.}

\figcaption[]{(a) A comparison between the \civ\ (solid line) and \mgii\
(dashed line) emission-line profiles as observed on 1996, February 21.
Each line has been transformed into velocity space, and normalized to
the total flux in a 2000~km~s$^{-1}$ wide bin centered at
$+$2000~km~s$^{-1}$. Note the good agreement between the line profile
shapes in the blue and red wings of the lines. The far blue wing of
\mgii\ is contaminated by blended \feii\ hence the poor agreement in
this region. (b) The high-state (epoch 2, solid line) and low-state
(epoch 5, dotted line) continuum subtracted \civ+\heii\ emission
lines, scaled to the integrated flux in a 2000~km~s$^{-1}$ wide bin
centered at $-$4000~km~s$^{-1}$.  The most significant shape changes
occur redward of line center, with the core displaying a deficit of
response, whereas the far red wing displays an enhanced response.  A
similar result is found for \lya.}


\newpage
 
\begin{references}

\reference{BOTT97}Bottorff, M., Korista, K.T., Shlosman, I., and Blandford, R.D. 1997,
ApJ 479, 200.

\reference{CAG98}Cagnoni, I., Fruscione, A., McHardy, I.M., and Papadakis, I.E. 1998, 
In : "Dal nano al Tera eV: tutti i colori degli AGN", third italian
conference on AGNs, Roma, May 18-21 1998, Memorie S.A.It.

\reference{CREN98}Crenshaw, D.M., Maran, S.P. and Mushotzky, R.F. 1998, ApJ 496, 797.

\reference{EBS92}Emmering, R.T., Blandford, R.D., and Shlosman, I. 1992, ApJ 385, 460.

\reference{GOAD93}Goad, M.R., O'Brien, P.T. and Gondhalekar,
 P.M. 1993, MNRAS 263, 149.

\reference{GOAD95}Goad, M.R. 1995, PhD Thesis, University College London.

\reference{GOAD98}Goad, M.R. et~al. 1999, ApJ submitted.

\reference{GK99}Goad, M.R. and Knigge, C. 1999, in prep.

\reference{KOLM93}Kolman, M. et~al. 1993, ApJ 403, 592.

\reference{AK97}Koratkar, A.P. et~al. 1996, ApJ 470, 378.

\reference{KK96}Krolik, J.H. and Kallman, T.R. 1988, ApJ 324, 714.

\reference{LA97}Laor, A., Jannuzi, B.T., Green, R.F., and Boroson,
 T.A. 1997, ApJ 489, 656.

\reference{MA97}Marshall, H.L. et~al. 1997, ApJ 479, 222.

\reference{MAT97}Mathur, S., Wilkes, B.J., and Aldcroft, T. 1997, ApJ 478, 182.

\reference{OB95}O'Brien, P.T. Goad, M.R. and Gondhalekar, P.M. 1995, MNRAS 275, 1125.

\reference{RO98}Romano, P. and Peterson, B.M. 1998, In :
"Structure and Kinematics of Quasar Broad Line Regions", Eds
C. M. Gaskell, W. N. Brandt, M. Dietrich, D. Dultzin-Hacyan, and
M. Eracleous, ASP Conf. Ser.

\reference{VOIT87}Voit, G.M., Shull, J.M., and Begelman, M.C. 1987, ApJ 316, 573.

\reference{WALT90}Walter, R., Ulrich, M.-H., Courvoisier, T.J.-L. and
 Buson, L.M. 1990, A\&A 233, 53.

\reference{WAND93}Wanders, I. et~al. 1993, A\&A 269, 39.

\reference{WAND95}Wanders, I. et~al. 1995, ApJL 453, L87.

\reference{WEYMANN91}Weymann, R.J., Morris, S.L., Foltz, C.B., Hewett, P.C. 1991,
 ApJ 373, 23.

\end{references}
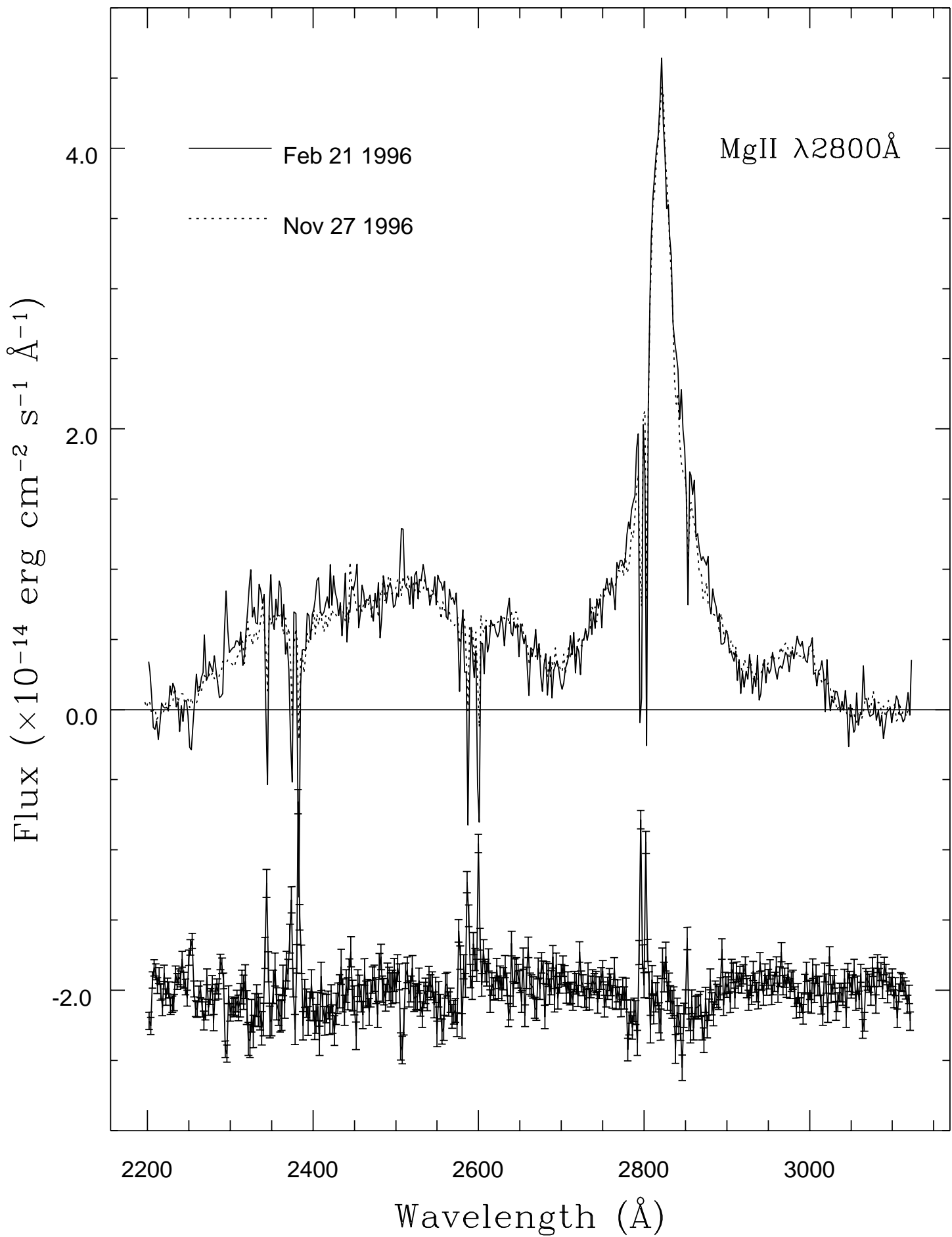
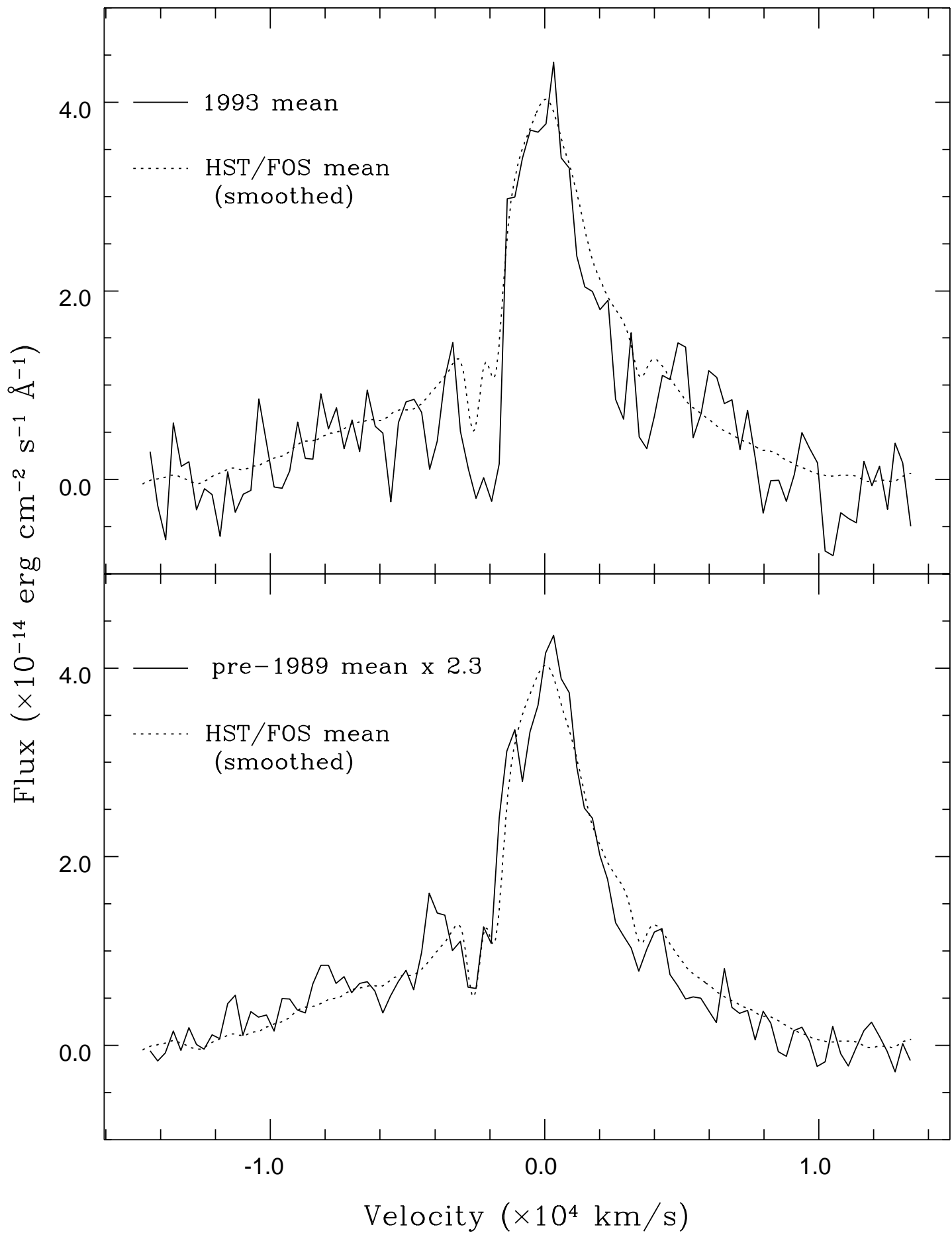
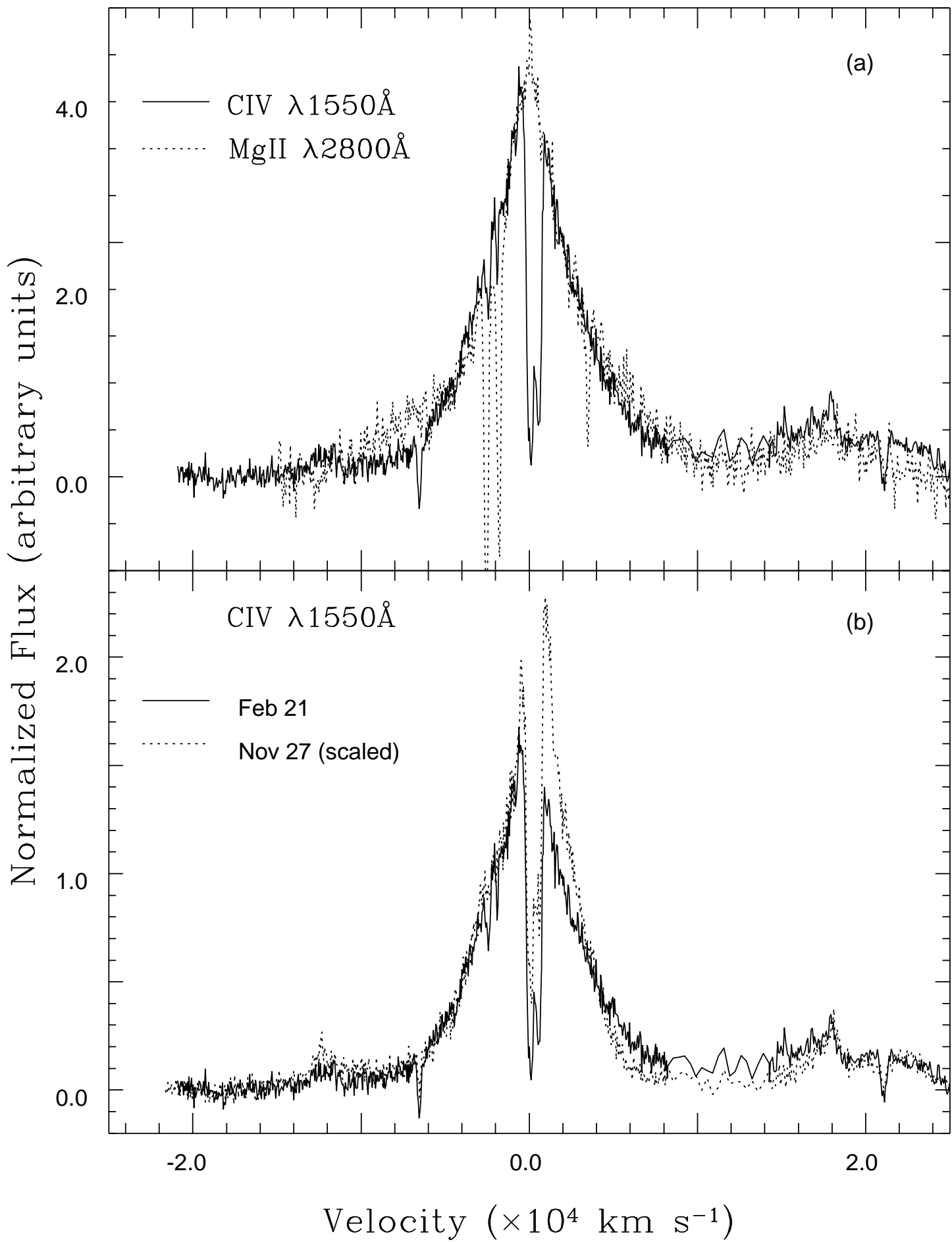
\end{document}